\documentclass[showpacs,preprint,pre]{revtex4}
\begin{document}
\begin{abstract}
Cilia are motile biological appendages that are driven to bend by internal shear stresses
between tubulin filaments.  A continuum model of ciliary material is constructed that
incorporates the essential ciliary constraints: (1) one-dimensional inextensibility
of filaments, (2) three-dimensional incompressibility, and (3) shear strain only along filaments.  This
hypothetical ciliary material combines one- and three-dimensional properties in a way that makes it
a natural and flexible model for how real cilia convert nanoscopic shear stress into motility
on a much larger scale.  Without reference to the evolving shape of the cilium,
conventional continuum mechanics applied to this hypothetical material
leads to the standard model of ciliary dynamics, but with one additional term, required by
constraints (2) and (3) above, a model-independent coupling of shear and twist in general ciliary motion.
\end{abstract}
\title{The Geometry of Ciliary Dynamics}
\author{Mark A. Peterson}
\affiliation{Mount Holyoke College}
\pacs{45.10.Na, 46.70.Hg, 47.63.Gd}
\email{mpeterso@mtholyoke.edu}
\date{\today}
\maketitle
\section{Introduction}
The beating of cilia, or eukaryotic flagella, has been the subject of modelling studies for over fifty years.
The hydrodynamics of propulsion is one aspect of these studies, concerned with Stokes flows
past the cilium at a mesoscale of tens of microns \cite{GrayHancock}\cite{Lighthill}\cite{BrokawJohnson}\cite{GueronLiron1992}
\cite{GueronLevitGurevich2001}.
This scale is set by the length of the cilium.
The mechanism of motility is another aspect, concerned with the
dynein mediated sliding of microtubule doublets on each other
at a nanoscale of tens of nanometers \cite{BrokawSliding}
\cite{Lubliner}\cite{HinesBlum1978}\cite{HinesBlum1983}\cite{HinesBlum1984}\cite{HinesBlum1985}.
This scale is set by the diameter of the cilium or even by the
size of the motor proteins.
It is the aim of this paper to bridge the gap between these scales.

The remarkable discrepancy between these length scales
poses a problem for modelling.  To put it simply, at the mesoscale the
cilium looks one-dimensional, but at the nanoscale it looks three-dimensional.
The discrepancy is most telling when the ciliary beat itself is three-dimensional.
In modelling the cilium at the mesoscale, the emphasis is on the exterior of the cilium regarded
as a space curve.  The terms of the description are dictated by this curve, whether in
the Frenet description of Gueron and Liron's three-dimensional ciliary beat model \cite{GueronLiron1993}, or
in the `body co-ordinates' of a later version by Gueron and Levit-Gurevich \cite{GueronLevitGurevich2001}.
The three-dimensional nanoscopic models of Hines
and Blum \cite{HinesBlum1978}\cite{HinesBlum1983}\cite{HinesBlum1984}\cite{HinesBlum1985},
on the other hand, use body coordinates truly anchored in the three-dimensional cilium,
and moving with it,
but these authors consider only static stress-strain relationships and do not address the
three-dimensional mesoscale dynamics
of the cilium.  Ultimately one would like to bridge the gap between these scales
so that observations of behavior at the mesoscale, which are increasingly accessible through high
speed video microscopy, could be simply related to geometry
at the nanoscale.

This article is essentially an abstraction from the sliding filament
models of Hines and Blum.  Although these authors work on detailed three-dimensional
structures at the nanoscale, they consistently summarize the results in effective parameters
for mesoscale models.  In this article I determine what must be true of all such models,
given only the constraints of sliding filament models in general.  It turns out
that the constraints at the nanoscale show up in dynamics at the mesoscale, and this is a model-independent
result.

The plan of the paper is as follows.  Section \ref{CiliaryMaterial} gives an abstract characterization
of a three dimensional material that I call `ciliary material,' subject to the constraints of the sliding
filament model:  it is incompressible, inextensible along one-dimensional ``filaments,"
and admits shear strain only longitudinally along the filaments.  This is an attempt to capture,
in a continuum model, the properties of the cilium.  These are the constraints consistently
invoked by Hines and Blum, but without reference to nanoscopic details.  These constraints tightly couple longitudinal
shear and bending strain, the basic mechanism of ciliary action.  Such ciliary material might
also model other biological structures, like auditory hair cells, where this coupling
has a different function.

Section \ref{CiliaryFlows}
determines how the ciliary material can move, subject to its constraints,
what I call ciliary flow.  Remarkably, although it
is a three-dimensional material, its motion is parameterized by the motion of any single
one-dimensional filament, and this one filament can move arbitrarily.  Thus the
hypothetical ciliary material, motivated by the structure of cilia,
combines, in an unexpected way, three-dimensional and one-dimensional properties in a manner that
is reminiscent of real cilia.

The transition in Section \ref{CiliaryDynamics}
from the nanoscopic properties of three-dimensional ciliary
material to the mesoscale dynamics
of an essentially one-dimensional cilium is a very short step.  The treatment there looks one dimensional, but
the description retains information about the three dimensional nanostructure because of the
peculiar properties of ciliary material.  The most significant result is a model-independent
coupling between shear and twist in three dimensional ciliary beats,
a phenomenon discussed by Hines and
Blum in specific models, but not recognized by them as a generally necessary feature of every
sliding filament model.

The geometrical methods in Sections \ref{CiliaryMaterial} and \ref{CiliaryFlows} are
not used very often in physics, but they are indispensable here, and I hope any reader who is unfamiliar
with them will find that this application is a useful introduction to them in a relevant context.
The books by Schutz \cite{schutz} and Frankel \cite{frankel} are good references.

\section{Ciliary Material}
\label{CiliaryMaterial}
In terms of smooth coordinates $(x^1,x^2,x^3)$ in space one can describe the deformation of any material
by the trajectories of its constituent particles, solutions of equations of motion
\begin{equation}
\label{EqOfMotion}
\frac{dx^i}{dt}=V^i(x,t)
\end{equation}
where
\begin{equation}
V=V^i\partial_i
\end{equation}
is the velocity vector field, and $t$ is time.
Metric relations among particles are given by
\begin{equation}
\label{ds2}
ds^2=g_{ij}dx^idx^j
\end{equation}
where $g_{ij}$ is the Euclidean metric.
Coordinatize the material object by Lagrangian coordinates
convected by the flow, i.e., let every material point keep the same coordinates that it had
originally.  The changing metric relationship of material points
is then expressed entirely by the change in the metric components $g_{ij}$, and
the rate of change
is given by the Lie derivative
\begin{equation}
\label{LVg}
\frac{\partial g_{jk}}{\partial t}=Vg_{jk}+g([\partial_j,V],\partial_k)+g(\partial_j,[\partial_k,V])
\end{equation}
Here $[~,~]$ is the Lie bracket of vector fields.

As an example of such a computation, let $V$ be the shearing flow
\begin{equation}
\label{VSimpleShear}
V=yS\partial_x
\end{equation}
in the plane, where $S$ is a constant.  Then if $x$ and $y$ are initially Cartesian coordinates,
the metric tensor in convected coordinates at time t is given by the Lie-Taylor series
(which terminates in this case)
\begin{equation}
g+t\pounds_Vg+\frac{t^2}{2!}\pounds_V\pounds_Vg=
\left(\begin{array}{cc}
  1 & tS \\
  tS & 1+t^2S^2
\end{array}\right)\,,
\end{equation}
and evaluating at $t=1$ one has the Euclidean metric
\begin{equation}
\label{g2}
g=\left(\begin{array}{cc}
  1 & S \\
  S & 1+S^2
\end{array}\right)\,,
\end{equation}
in skew coordinates
corresponding to constant shear strain $S$.

The flow $V$ of Eq.~(\ref{VSimpleShear}) is an example of a ciliary flow, and the
form of the metric tensor $g$ in Eq.~(\ref{g2}) encodes the result.  The
filaments, parallel to the $x$-axis and labeled by the $y$ coordinate,
slide on each other inextensibly, as one sees in the
metric component $g_{11}=1$.  The resulting shear strain is visible in the off-diagonal component $g_{12}=S$.
Finally, ${\rm det}\,g=1$ means that the flow is incompressible.  The three dimensional
generalization of this form to a general ciliary configuration is
\begin{equation}
\label{g3}
g=\left(\begin{array}{ccc}
  1 & S & T\\
  S & 1+S^2 & ST\\
  T & ST & 1+T^2
\end{array}\right)\,.
\end{equation}
where now $S$ and $T$, two independent components of shear strain, depend on Lagrangian coordinates $(x,y,z)$.
Here $x$ is arclength along filaments, up to an additive constant reflecting the arbitrariness
of choosing an origin in each filament, and the coordinates $(y,z)$ label the filaments.  It should be
noted that this is not the most general outcome of an incompressible flow of the filaments, because
the only motion that has been allowed to them is sliding longitudinally along each other (and of course bending).
These are exactly the constraints
of the sliding filament models of Hines and Blum, encoded in the form of $g$.

Associated with the metric tensor $g$ is an orthonormal frame field
\begin{eqnarray}
\label{e1}
e_1&=&\partial_x\\
\label{e2}
e_2&=&\partial_y-S\partial_x\\
\label{e3}
e_3&=&\partial_z-T\partial_x\,.
\end{eqnarray}
where $e_1$ is everywhere tangent to filaments.  This is not, however, a coordinate frame, because
the vector fields do not commute as differential operators, in general.

The metric $g$ determines the shape of each filament up to a rigid motion, because for each $(y,z)$ it determines
the filament's curvature $\kappa$ and torsion $\tau$ as a function of $x$ (see Appendix \ref{SpaceCurves}).  These are the
Frenet data for the filament as a space curve.
One has
\begin{eqnarray}
\label{kappaST}
\kappa&=&\sqrt{S_x^2+T_x^2}\\
\label{tauST}
\tau&=&\frac{T_{xx}S_x-S_{xx}T_x}{\kappa^2}
\end{eqnarray}
where the subscript $x$ denotes the derivative $\partial_x$ along the filament.
The Frenet equations for the filaments then determine the space curves $\vec{R}(x)$ and an
orthonormal frame $\{\hat{T},\hat{N},\hat{B}\}$ at each point of each curve according to
\begin{eqnarray}
\label{Frenet}
\vec{R}_x&=&\hat{T}\\
\hat{T}_x&=&\kappa\,\hat{N}\\
\hat{N}_x&=&-\kappa\,\hat{T}+\tau\hat{B}\\
\hat{B}_x&=&-\tau\,\hat{N}
\end{eqnarray}
In the context of ciliary geometry it is more natural to describe the filaments in terms of
the shear strains $S$ and $T$ directly
than to translate them into the Frenet language.  This is an alternative, equivalent description
of space curves.  It is even possible to translate the other way, from the Frenet description to the
ciliary description, inverting the relations in Eqs.~(\ref{kappaST})-(\ref{tauST}),
\begin{eqnarray}
\label{Sprime}
S_x&=&\kappa\cos\phi\\
\label{Tprime}
T_x&=&\kappa\sin\phi\\
\phi_x&=&\tau
\end{eqnarray}
In this formulation the space curves $\vec{R}(x)$ are the solutions of
\begin{eqnarray}
\label{Rprime}
\vec{R}_x&=&e_1\\
\label{e1prime}
\partial_x e_1&=&\kappa\cos\phi\, e_2+\kappa\sin\phi\, e_3\\
\label{e2prime}
\partial_x e_2&=&-\kappa\cos\phi\, e_1\\
\label{e3prime}
\partial_x e_3&=&-\kappa\sin\phi\, e_1\\
\label{phiprime}
\phi_x&=&\tau
\end{eqnarray}
There is an arbitrary constant in the angle $\phi$ since, unlike
$\hat{N}$ and $\hat{B}$ in the Frenet picture, $e_2$ and $e_3$ are only determined up to a global rotation about $e_1$.
This amounts to a gauge freedom in the ciliary description.  An advantage of the ciliary description
is that it remains well defined where the curvature vanishes,
a nuisance in the Frenet description.

The shear strains $S$ and $T$ cannot be arbitrary functions, because $g$ is the Euclidean
metric, even if it is expressed in peculiar coordinates.  Its associated Riemannian curvature
tensor must therefore vanish identically, and hence $S$ and $T$ must obey the following
identities (see Appendix \ref{FlatGeometry})
\begin{eqnarray}
\label{SSx}
0&=& \partial_x(e_2 S)= \partial_x(e_3 S)=\partial_x( e_2 T)= \partial_x(e_3 T) \\
\label{e2e3integ}
0&=&e_3 S-e_2 T\,.
\end{eqnarray}
The second of these equations can be recognized as
\begin{equation}
\label{integ}
[e_2,e_3]=0\,.
\end{equation}
This says that $e_2$ and $e_3$ together form an integrable distribution, and thus there exist surfaces
normal to the filaments, at least locally.
Moreover, since $e_2$ and $e_3$ are now an orthonormal coordinate frame field
on these surfaces, they are Euclidean planes, which I will call normal planes.

By allowing the filaments to reptate along their length inextensibly,
one can bring Eq.~(\ref{SSx})
to the simpler form
\begin{equation}
\label{SSxsimple}
0=e_2 S=e_3 S=e_2 T=e_3 T\,
\end{equation}
that is, the strains $S$ and $T$ can be made {\it constant} in normal planes.  Thus the configuration of
filaments in a neighborhood of a given filament is entirely determined by that filament.  The
given filament determines the direction of neighboring filaments, since it determines the
normal planes, and it determines their curvatures and torsions, since it determines the values of
$S$ and $T$.  The geometrical meaning of Eqs.~(\ref{SSx})-(\ref{e2e3integ}) is that these two
potentially conflicting descriptions are consistent.  Thus the ciliary geometry is almost rigid.
The freedom that it possesses corresponds essentially to a single free space curve.

I turn now to the motions possible in ciliary matter, ciliary flows.  Since one
of those motions is reptation of filaments, I will also justify the assertions of the preceding
paragraph.

\section{Ciliary Flows}
\label{CiliaryFlows}
A ciliary flow, by definition, must maintain the form of Eq.~(\ref{g3}) even as $g$ evolves according
to Eq.~(\ref{LVg}).  If $V$ is the flow with components $(\alpha,\beta,\gamma)$, namely
\begin{equation}
\label{Vabc}
V=\alpha\partial_x+\beta\partial_y+\gamma\partial_z\,,
\end{equation}
then for $V$ to be ciliary $(\alpha,\beta,\gamma)$ must obey the conditions (see Appendix \ref{AppCiliaryFlows})
\begin{eqnarray}
\label{div0}
0&=&\alpha_x+S\beta_x+T\gamma_x\\
\label{e2b}
0&=&e_2\beta\\
\label{e3c}
0&=&e_3\gamma\\
\label{e2ce3b}
0&=&e_2\gamma+e_3\beta\\
\label{twistprime}
0&=&(e_2\gamma)_x-\beta_xT_x+\gamma_xS_x\\
\label{e2e2gamma}
0&=&e_2e_2\gamma\\
\label{e3e3beta}
0&=&e_3e_3\beta
\end{eqnarray}
where the partial derivative $\partial_x$ is indicated by the subscript $x$.
Eqs.~(\ref{e2e2gamma})-(\ref{e3e3beta}) are not independent of the others, since
for example $e_3e_3\beta=-e_3e_2\gamma=-e_2e_3\gamma=0$.  In fact all higher derivatives
of $\beta$ and $\gamma$ in the normal planes vanish by this argument, so that $\beta$
and $\gamma$ restricted to a normal plane can only be affine linear functions.
I return to this consideration below.

The simplest nontrivial ciliary flow is reptation, $\alpha_x=\beta=\gamma=0.$
By Eq.~(\ref{LVg}) the shear strains change under this flow at the rate
\begin{eqnarray}
\label{ReptationSt}
S_t&=&\alpha_y+\alpha S_x\\
\label{ReptationTt}
T_t&=&\alpha_z+\alpha T_x\,,
\end{eqnarray}
where subscripts indicate partial derivatives with respect to the corresponding variable.
It is possible to construct a reptation $\alpha$ that alters $S$ and $T$ in such 
a way that Eq.~(\ref{SSxsimple}) holds
in one normal plane, at least locally (see Appendix \ref{AppOnePlane}).  Then by Eq.~(\ref{SSx}) $S$ and $T$ are constant in
every normal plane along the filaments.  This proves the assertions made at the
end of the last section.  I will now assume that $S$ and $T$ obey the simpler
Eq.~(\ref{SSxsimple}), since this can
always be arranged by a reptation.

The ciliary conditions require $\beta$ and $\gamma$ to be affine
linear functions in the normal planes.  To see that there exist non-trivial
solutions to these conditions, imagine specifying $\beta(x)$ and $\gamma(x)$
arbitrarily along one filament.  Integrating Eq.~(\ref{twistprime}) determines $e_2\gamma=-e_3\beta$
along the filament up to an arbitrary global constant.  These are all the data required
to extend $\beta$ and $\gamma$ in each normal plane as affine
linear functions.  This extension
continues to satisfy Eq.~(\ref{twistprime}) on neighboring filaments because of the
easily verified identities
\begin{eqnarray}
e_2[(e_2\gamma)_x-T_x\beta_x+S_x\gamma_x]&=&2S_x[(e_2\gamma)_x-T_x\beta_x+S_x\gamma_x]\\
e_3[(e_2\gamma)_x-T_x\beta_x+S_x\gamma_x]&=&2T_x[(e_2\gamma)_x-T_x\beta_x+S_x\gamma_x]
\end{eqnarray}
Repeated differentiation of Eq.~(\ref{twistprime}) using $e_2$ and $e_3$ shows that
if $(e_2\gamma)_x-T_x\beta_x+S_x\gamma_x$ vanishes on a filament,
then all its normal derivatives also vanish there.
Thus given only that it is represented by its Taylor series, it is constant, and hence zero,
and the constructed solution obeys all the ciliary conditions.
$V$ is a non-trivial ciliary flow, determined by its values
on one filament.

The conditions Eqs.~(\ref{div0})-(\ref{e3e3beta}) confirm that the ciliary configuration is
entirely determined by one filament.  If one tries to move the filaments, one can specify $\beta(x)$
and $\gamma(x)$ on only one given filament.  The ciliary conditions then determine $\alpha(x)$ along that
filament up to a constant (a reptation).
They further determine $\beta$ and $\gamma$ almost uniquely as linear functions in normal
planes.  Neighboring filaments intersect the normal planes (each plane labelled by $x$, its intersection
with the given filament), and
in the course of the motion these intersection points
rotate about the given filament at
an angular velocity
\begin{equation}
\label{omega}
\omega(x)=(e_2\gamma)(x)
\end{equation}
which is not arbitrary but is determined up to a global constant by $\beta$ and $\gamma$ according to
Eq.~(\ref{twistprime}).  (The undetermined constant in $\omega$ describes a global twisting rotation in which every normal
plane rotates at the same rate, a motion which is possible but probably not
physically relevant.)
The remaining freedom in $\alpha$ corresponds to reptations of neighboring filaments,
also probably not physically relevant.
Thus, to summarize, any filament determines the configurations of its neighbors, and any motion
of that filament determines the motion of its neighbors.
It is worth noting what was not obvious {\it a priori}, that non-trivial motions of the filaments are
possible, that is, the ciliary material is not completely rigid.  One filament can move arbitrarily
(but inextensibly),
and all the others must follow it.

\section{Ciliary Dynamics}
\label{CiliaryDynamics}
The surprisingly one-dimensional character of ciliary matter means that its restriction
to modelling the cilium is immediate.  Represent the cilium as a thin cylinder of
ciliary matter, with length $L$ and radius $\rho$, all dynamical quantities now depending just
on $x$, the coordinate along one given filament.  Let the cilium have
elastic moduli, so
that the energy of a configuration is given by the shear energy and bending energy
\begin{equation}
E=\frac{\mu}{2}\int_0^L [(S-F)^2+(T-G)^2]\,dx+\frac{\kappa_c}{2}\int_0^L(S_x^2+T_x^2)\,dx\,,
\end{equation}
where $\mu$ is the shear modulus and $\kappa_c$ is the bending modulus (with dimensions
appropriate to one dimension, not three).  Here $F$ and $G$
are target shear strains, such that the shear energy would be minimal if the shear strains
$S$ and $T$ could relax to $F$ and $G$.  The operation of the ciliary ``engine" would be to
make $F$ and $G$ functions of time, so that the equilibrium becomes a moving target.
Asymmetries in the nanoscopic structure of the cilium could be built into this
expression, which is taken symmetrical here for illustration.

Conjugate to shear strains $S$ and $T$ are internal stresses
\begin{eqnarray}
\Phi&=&\delta E/\delta S=\mu(S-F)-\kappa_c S_{xx}\\
\Psi&=&\delta E/\delta T=\mu(T-G)-\kappa_c T_{xx}\,
\end{eqnarray}
which, by the above mentioned one-dimensionality, have the units of force.

Continuing this variational approach to the dynamics, one finds the generalized forces conjugate
to the displacement of a cilium along $V=\alpha\partial_x+\beta\partial_y+\gamma\partial_z$ of Eq.~(\ref{Vabc}),
together with the required rotation $\omega$ of Eq.(\ref{omega}) in normal planes.
In the orthonormal basis, $V$ takes the form
\begin{equation}
V=(\alpha+\beta S+\gamma T)e_1+\beta e_2+\gamma e_3\,,
\end{equation}
a representation that becomes increasingly appropriate as one moves up to the mesoscopic scale.
One notes that $\beta$ and $\gamma$ are the components of velocity normal to the cilium,
but that they also contribute to the tangential velocity if the shear strains are non-zero.
This is a residual piece of three dimensional information in the one-dimensional
description.

From Eq.~(\ref{LVg}), imposing Eqs.~(\ref{div0})-(\ref{e3e3beta}),
the strains under the flow $V$ change at the rate
\begin{eqnarray}
\label{St}
S_t&=& (\alpha +\beta S+\gamma T)S_x+\beta_x+e_2\alpha + \omega T\\
\label{Tt}
T_t&=&(\alpha +\beta S +\gamma T) T_x+\gamma_x+e_3\alpha-\omega S
\end{eqnarray}
The terms $e_2\alpha$ and $e_3\alpha$ represent reptations within the cilium,
a possibility that is ignored
from now on.
The constraints
\begin{eqnarray}
\label{constraintlambda}
0&=&\alpha_x+S\beta_x+T\gamma_x\\
\label{constraintnu}
0&=&\omega_x-T_x\beta_x+S_x\gamma_x
\end{eqnarray}
are handled with Lagrange multipliers $\lambda$ and $\nu$ in the expression
\begin{equation}
E'=E+\int_0^L \lambda(\alpha_x+S\beta_x+T\gamma_x)\,dx+\int_0^L\nu(\omega_x-T_x\beta_x+S_x\gamma_x)\,dx\
\end{equation}
Then the generalized interior forces on the cilium are
\begin{eqnarray}
F_\alpha=-\delta E'/\delta\alpha&=&-\Phi S_x-\Psi T_x+\lambda_x\\
F_\beta=-\delta E'/\delta\beta&=&-\Phi S S_x-\Psi S T_x+\Phi_x+(\lambda S)_x-(\nu T_x)_x\\
F_\gamma=-\delta E'/\delta\gamma&=&-\Phi T S_x-\Psi T T_x+\Psi_x+(\lambda T)_x+(\nu S_x)_x\\
F_\omega=-\delta E'/\delta\omega&=&\nu_x-\Phi T+\Psi S
\end{eqnarray}
Generalized external forces on the cilium, due to the fluid medium of viscosity $\eta$ in which it is immersed,
can be found from the dissipation function $D$ for the
corresponding Stokes flow.  A thorough analysis of this problem has been done by Gueron
and Liron \cite{GueronLiron1992}.  For simplicity I take just the leading term indicated by their analysis,
\begin{equation}
\label{dissipationD}
D=\int_0^L\left[\frac{C_T}{2}(\alpha+\beta S+\gamma T)^2+\frac{C_N}{2}(\beta^2+\gamma^2)+\frac{C_\omega}{2}\omega^2\right]\,dx
\end{equation}
where the C's are constant resistance coefficients.
%, with $C_N/C_T\approx 1.8$ and $C_\omega\approx 4\pi\eta \rho^2$,
%with $\rho$ the radius of the cilium.
The viscous force conjugate to displacement by $\alpha$ is then $-\delta D/\delta\alpha$, etc.  Requiring that
elastic forces and viscous forces balance, i.e.,
\begin{equation}
\frac{\delta E'}{\delta\alpha}=\frac{\delta D}{\delta\alpha}\,,
\end{equation}
etc., leads to the surprisingly simple dynamical laws connecting the motion of ciliary matter to
its internal stresses,
\begin{eqnarray}
\label{alphasol}
C_T(\alpha+\beta S+\gamma T)&=&-\Phi S_x-\Psi T_x + \lambda_x\\
\label{betasol}
C_N\beta&=&\Phi_x+\lambda S_x-(\nu T_x)_x\\
\label{gammasol}
C_N\gamma&=&\Psi_x+\lambda T_x+(\nu S_x)_x\\
\label{omegasol}
C_\omega\omega&=&\nu_x-\Phi T+\Psi S
\end{eqnarray}
Under this flow the strains change according to Eqs.~(\ref{St})-(\ref{Tt}).
The Lagrange multipliers can now be determined from the constraints
Eqs.~(\ref{constraintlambda})-(\ref{constraintnu}).
Substituting the solutions in Eqs.~(\ref{alphasol})-(\ref{omegasol}) gives the equations
for $\lambda$ and $\nu$,
\begin{eqnarray}
\label{lambdaequation}
\lambda_{xx}&=&\frac{C_T}{C_N}\kappa^2\lambda+(\Phi S_x+\Psi T_x)_x+\frac{C_T}{C_N}(S_x\Phi_x+T_x\Psi_x)
+\frac{C_T}{C_N}\tau\kappa^2\nu\\
\label{nuequation}
\nu_{xx}&=&(\Phi T-\Psi S)_x-\frac{C_\omega}{C_N}\left[S_x\Psi_{xx}-T_x\Phi_{xx}
+\lambda\tau\kappa^2+T_x(\nu T_x)_{xx}+S_x(\nu S_x)_{xx}\right]\,,
\end{eqnarray}
where I have used the expressions for curvature $\kappa$ and torsion $\tau$ first derived in terms of $S$ and $T$
 in Eqs.~(\ref{kappaST})-(\ref{tauST}).
In the limit as $C_\omega/C_N\rightarrow0$, the equation for $\nu$ simplifies considerably,
with solution
\begin{equation}
\label{nusimple}
\nu=\int_0^x (\Phi T-\Psi S)\,dx + c_1 x+c_2
\end{equation}
The solution for $\omega$ is already explicit in terms of $\beta$ and $\gamma$
from the constraint Eq.~(\ref{constraintnu}),
\begin{equation}
\label{omegasimple}
\omega=\int_0^x (T_x\beta_x-S_x\gamma_x)\,dx+{\rm const.}
\end{equation}

With one important exception, the above dynamical laws for ciliary matter are
precisely the usual phenomenological laws for three-dimensional motions of cilia, as derived by
Gueron and Liron \cite{GueronLiron1993}.  It is remarkable that they emerge here without any
appeal or reference to the shape of the cilium (apart from the hydrodynamic interaction
in a more accurate dissipation function, Eq.~(\ref{dissipationD})), 
but purely as a consequence of nanoscopic processes.  The
occurrence of $\kappa$ and $\tau$ above is just an abbreviation for certain combinations of
derivatives of $S$ and $T$ that appeared.  Of course the large scale shape of the cilium could be reconstructed at any
time from Eqs.~(\ref{Sprime})-(\ref{phiprime}).

The dynamical laws have been
expressed here in terms of the force vector
\begin{equation}
\vec{F}=\lambda e_1+\Phi e_2 + \Psi e_3
\end{equation}
but the vector quantities in \cite{GueronLiron1993} were expressed in terms of the Frenet basis $\hat{T},\hat{N},\hat{B}$,
where $\vec{F}$ took the form
\begin{equation}
\vec{F}=-F_T\hat{T}-F_N\hat{N}-F_B\hat{B}\,.
\end{equation}
From Eqs.~(\ref{Sprime})-(\ref{e1prime}) one has the transformation that
connects these two descriptions,
\begin{eqnarray}
\hat{T}&=&e_1\\
\hat{N}&=&\cos\phi\, e_2+\sin\phi\, e_3\\
\hat{B}&=&-\sin\phi\, e_2+\cos\phi\, e_3
\end{eqnarray}
with $\phi_x=\tau$, the torsion of the curve.  Using also Eqs.~(\ref{kappaST})-(\ref{tauST}),
it is straightforward to verify that the dynamical laws of ciliary matter, as derived here, agree
with standard phenomenology in every respect but one:
in ciliary matter there is an additional constraint, and a corresponding additional
term in the laws.

The additional constraint arises because the sliding filament model is constrained by
more than just the inextensibility of the filaments.  Its constraints are expressed in
the form of the strain tensor, or equivalently in the form of the metric tensor in
Lagrangian coordinates, Eq.~(\ref{g3}).  It is to be expected that these additional
constraints would have consequences for the dynamics, just as inextensibility does,
through the Lagrange multiplier $\lambda$.  The dynamical form
that this constraint takes is a certain definite coupling of shear and twist, the
last terms in Eqs.~(\ref{betasol})-(\ref{gammasol}), and more generally every reference
to the Lagrange multiplier $\nu$ and the twist velocity $\omega$.

Hines and Blum, in developing detailed nanoscopic models within the framework
of the sliding filament model, noticed the possibility of such a coupling and wrote more than one paper
about it \cite{HinesBlum1984}\cite{HinesBlum1985}, pointing out also experimental observations of the effect.
It might have seemed to their readers, however, that this coupling depended upon nanoscopic details,
and that there was no necessity for it.  No one seems to have pointed out
that the coupling of shear and twist is a {\it model independent} consequence
of the sliding filament constraints, and must be included in any valid sliding filament model,
including dynamical models of behavior at the mesoscale.

\begin{appendix}
\section{Ciliary filaments as space curves}
\label{SpaceCurves}
The rates of change of the vector fields $e_2$, $e_3$, along the integral curves of $e_1$, (the filaments
of the ciliary space) are
given by the Lie derivatives (Lie brackets)
\begin{eqnarray}
\partial_x e_2&=&[e_1,e_2]=-S_x e_1\\
\partial_x e_3&=&[e_1,e_3]=-T_x e_1
\end{eqnarray}
where the subscript $x$ indicates $\partial_x$, the derivative along a filament.
Thus the curvature vector, in the sense of the Frenet equations, is
\begin{equation}
\kappa \hat{N}=S_x e_2+T_x e_3
\end{equation}
where the curvature is the norm of this vector
\begin{equation}
\kappa=\sqrt{S_x^2+T_x^2}\,.
\end{equation}
The rate of change of the principal normal $\hat{N}$ along the filament, dotted with the binormal
\begin{equation}
\hat{B}=\frac{S_xe_3-T_xe_2}{\kappa}
\end{equation}
is the torsion
\begin{equation}
\tau=g([e_1,\hat{N}],\hat{B})=\frac{T_{xx}S_x-S_{xx}T_{x}}{\kappa^2}
\end{equation}

\section{Consequences of flat geometry}
\label{FlatGeometry}
Differential forms dual to the orthonormal vector fields $e_1$, $e_2$, $e_3$ of Eqs.~(\ref{e1})-(\ref{e3})
are
\begin{eqnarray}
\sigma^1&=&dx+Sdy+Tdz\\
\sigma^2&=&dy\\
\sigma^3&=&dz\,.
\end{eqnarray}
Define
\begin{equation}
A=\frac{1}{2}(TS_x-ST_x+T_y-S_z)\,.
\end{equation}
It is straightforward to verify that the connection forms
\begin{eqnarray}
\omega^1_{~1}&=&\omega^2_{~2}=\omega^3_{~3}=0\\
\omega^1_{~2}&=&-\omega^2_{~1}=-S_x\sigma^1+A\sigma^3\\
\omega^1_{~3}&=&-\omega^3_{~1}=-T_x\sigma^1-A\sigma^2\\
\omega^2_{~3}&=&-\omega^3_{~2}=-A\sigma^1
\end{eqnarray}
satisfy the Cartan structure equations
\begin{equation}
\label{CartanStructure}
d\sigma^i+\omega^i_{~j}\wedge\sigma^j=0\,.
\end{equation}
Now because the metric is flat, the curvature forms
\begin{equation}
\theta^i_{~j}=d\omega^i_{~j}+\omega^i_{~k}\wedge\omega^k_{~i}
\end{equation}
must vanish identically.  In particular,
\begin{equation}
\theta^2_{~3}=-2A\,\sigma^2\wedge\sigma^3+ {\rm other~components}
\end{equation}
so $A$ must vanish, and this is Eq.~(\ref{e2e3integ}).  The remaining identities reduce to $d\omega^1_{~2}=d\omega^1_{~3}=0$,
and this is Eq~(\ref{SSx}).

\section{Geometrical calculations for ciliary flows}
\label{AppCiliaryFlows}
Under a ciliary flow $V$ it is necessary that $g_{11}$ keep the constant value 1,
so that
\begin{equation}
0=\partial_tg_{11}=2g([\partial_x,V],\partial_x)=2(\alpha_x+S\beta_x+T\gamma_x)\,,
\end{equation}
and this is Eq.~(\ref{div0}).  To preserve the form of $g$ it is necessary that
\begin{eqnarray}
\partial_t g_{22}&=&2SS_t=2S\partial_t g_{12}\\
\partial_t g_{33}&=&2TT_t=2T\partial_t g_{13}\\
\partial_t g_{23}&=&ST_t+TS_t=S\partial_t g_{13}+T\partial_t g_{12}
\end{eqnarray}
and these conditions, using Eq.~(\ref{LVg}), are Eqs.~(\ref{e2b})-(\ref{e2ce3b}).
Finally, it is not enough that $g$ keep the form of Eq.~(\ref{g3}), it must also continue to have zero
Riemannian curvature.  The most efficient way to handle this computation is to use the
moving orthonormal frame associated with the changing $g$.  The change in these vector fields
is computed by the Lie derivative (the Lie bracket for vector fields).  Thus under the ciliary flow $V$
of Eq.~(\ref{Vabc})
\begin{eqnarray}
\partial_t e_1&=&[e_1,V]=\alpha_xe_1+\beta_x\partial_y+\gamma_x\partial_z=\beta_xe_2+\gamma_x e_3\\
\partial_t e_2&=&[e_2,V]-(\partial_t S)e_1=\left(\sigma^1([e_2,V])-\partial_t S\right)e_1+(e_2\gamma)e_3\\
\partial_t e_3&=&[e_3,V]-(\partial_t T)e_1=\left(\sigma^1([e_3,V])-\partial_t T\right)e_1+(e_3\beta)e_2
\end{eqnarray}
As a check, one verifies
what is
required by orthonormality,
\begin{eqnarray}
\partial_t S&=&\sigma^1([e_2,V])+\beta_x\\
\partial_t T&=&\sigma^1([e_3,V])+\gamma_x\,,
\end{eqnarray}
using Eq.~(\ref{LVg}), and also $e_3\beta=-e_2\gamma$ by Eq.~(\ref{e2ce3b}).
To summarize,
\begin{eqnarray}
\label{e1rotate}
\partial_t e_{1}&=&\beta_xe_2+\gamma_x e_3\\
\label{e2rotate}
\partial_t e_{2}&=&-\beta_xe_1-(e_3\beta)e_3\\
\label{e3rotate}
\partial_t e_{3}&=&-\gamma_xe_1-(e_2\gamma)e_2
\end{eqnarray}
For the Riemannian curvature to vanish it is necessary that
\begin{eqnarray}
0&=&\partial_t[e_2,e_3]=[\partial_t e_2,e_3]+[e_2,\partial_t e_3]\\
 &=&[(e_3\beta)_x-(e_2\gamma)_x+2T_x\beta_x-2S_x\gamma_x]e_1+(e_2e_2\gamma)e_2+(e_3e_3\beta)e_3
 \end{eqnarray}
The vanishing of this vector field is conditions Eqs.~(\ref{twistprime})-(\ref{e3e3beta}).
It remains to show that with these conditions the other components of the Riemannian curvature
continue to vanish identically.
Dual to Eqs.~(\ref{e1rotate})-(\ref{e3rotate}) one has
\begin{eqnarray}
\partial_t \sigma^1&=&\beta_x\sigma^2+\gamma_x\sigma^3\\
\partial_t \sigma^2&=&-\beta_x\sigma^1-(e_3\beta)\sigma^3\\
\partial_t \sigma^3&=&-\gamma_x\sigma^1-(e_2\gamma)\sigma^2
\end{eqnarray}
The Cartan structure equations, which are identities, require
\begin{eqnarray}
\partial_t \omega^1_{~2}&=&(-\beta_{xx}+T_x(e_3\beta))\sigma^1-S_x\beta_x\sigma^2-S_x\gamma_x\sigma^3\\
\partial_t \omega^1_{~3}&=&(-\gamma_{xx}+S_x(e_2\gamma))\sigma^1-T_x\beta_x\sigma^2-T_x\gamma_x\sigma^3\\
\partial_t \omega^2_{~3}&=&0
\end{eqnarray}
and a long but straightforward computation then shows that
\begin{equation}
\partial_t(d\omega^i_{~j}+\omega^i_{~k}\wedge\omega^k_{~j})=0
\end{equation}
for all components.

\section{Making $S$ and $T$ constant in one normal plane by reptation}
\label{AppOnePlane}
Let $\alpha$ be a reptation, i.e., $V=\alpha\partial_x$ and $\alpha_x=0$.
Then, rearranging Eqs.~(\ref{ReptationSt})-(\ref{ReptationTt}), the convective derivative
\begin{eqnarray}
S_t-\alpha S_x&=&\alpha_y\\
T_t-\alpha T_x&=&\alpha_z
\end{eqnarray}
is the rate of change of strain in a fixed normal plane. (The normal planes are not
convected with the material but are associated with the stationary pattern of the filaments,
so to stay in a normal plane, subtract the convective term).
Thus in a fixed normal plane, since $\alpha_x=0$,
\begin{eqnarray}
(e_2S)_{t'}&=&e_2(S_t-\alpha S_x)=e_2\alpha_{y}=e_2e_2\alpha\\
(e_3T)_{t'}&=&e_3(T_t-\alpha T_x)=e_3\alpha_{z}=e_3e_3\alpha\\
(e_3S)_{t'}&=&(e_2 T)_{t'}=e_2e_3\alpha
\end{eqnarray}
Here the subscript $t'$ means the time derivative at a fixed normal plane, not at fixed $x$.
In this normal plane, we can introduce
coordinates $(y',z')$ such that $e_2=\partial/\partial y'$, $e_3=\partial/\partial z'$.
Then by the Poincar\'e lemma, Eq.~(\ref{e2e3integ}) says that there exists locally a function $F(y',z',t)$ such that in that plane
$S=e_2F$, $T=e_3F$.  Extend $F$ to be constant along filaments, and take $\alpha=-F$.
Then under the reptation $V=\alpha\partial_x$,
\begin{eqnarray}
(e_2S)_{t'}&=&-e_2S\\
(e_3T)_{t'}&=&-e_3T\\
(e_3S)_{t'}&=&(e_2T)_{t'}=-e_3S=-e_2T\,.
\end{eqnarray}
Thus these quantities decay to zero exponentially with time.  In the limit as $t\rightarrow\infty$, Eq.~(\ref{SSxsimple})
holds in one normal plane (and then by Eq.~(\ref{SSx}) it holds in all normal planes).
\end{appendix}

\end{document}